\documentclass{cjpsuppl}
\usepackage{graphicx}

\newcommand{\rar}{\rightarrow}
\newcommand{\pdup}{p_\uparrow}
\newcommand{\pimp}{\pi^- + \pdup \rar \pi^0 + X}
\newcommand{\ppdup}{p + \pdup \rar \pi^0 + X}

\begin{document}
\title{Single-Spin Asymmetry in 
Inclusive {\boldmath $\pi^0$}~Production Measured at the 
Protvino 70~GeV Accelerator.}
\authori{A.\,M. Davidenko, V.N.~Grishin,  V.Yu.~Khodyrev, V.I.~Kravtsov,
Yu.A.~Matulenko, V.A.~Medvedev, Yu.M.~Melnick, A.P.~Meschanin, 
V.V.~Mochalov, D.A.~Morozov, \underline {L.V.~Nogach}, S.B.~Nurushev, 
P.A.~Semenov, K.E.~Shestermanov, L.F.~Soloviev, A.N.~Vasiliev, A.E.~Yakutin}
\addressi{Institute of High Energy Physics, Protvino, Russia }
\authorii{N.S.~Borisov, A.N.~Fedorov, V.N.~Matafonov, A.B.~Neganov, 
Yu.A.~Plis, Yu.A.~Usov}    
\addressii{Joint Institute for Nuclear Research, Dubna, Russia}
\authoriii{A.A.~Lukhanin}   
\addressiii{Kharkov Physical Technical Institute, Kharkov, Ukraine}
\authoriv{}    \addressiv{}
\authorv{}     \addressv{}
\authorvi{}    \addressvi{}
\headtitle{Single-Spin Asymmetry Measurements at Protvino 70~GeV Accelerator}
\headauthor{L.V. Nogach }
\lastevenhead{L.V. Nogach: Single-Spin Asymmetry Measurements at Protvino Accelerator}
\refnum{}
\daterec{20 October 2003}
\suppl{A}  \year{2003} \setcounter{page}{1}
\maketitle

\begin{abstract}
Single Spin Asymmetries (SSA) $A_N$ measured in the two reactions 
at the Protvino 70~GeV accelerator are presented. $A_N$ in the reaction 
$\ppdup$ in the central region is close to zero within the error bars. 
SSA in the reaction $\pimp$ in the polarized target fragmentation region
is equal to $(-15 \pm 4)\%$ at $|x_F|>0.4$. There is an indication 
that the asymmetry arises at the same pion energy in the 
center of mass system. 

\end{abstract}

\section*{Introduction}

Large polarization effects were found during last few decades.
SSA was observed to be of order 
20--40\%,  while Perturbative Quantum Chromodynamics (pQCD) makes 
a qualitative prediction that the single-spin transverse effects 
should be very small due to the helicity conservation \cite{kane}.  
Here we present new SSA measurements carried out at the 
70~GeV Protvino accelerator in the reaction $\ppdup$ at 70~GeV in 
the central region ($x_F \approx 0$) and in the reaction $\pimp$ 
at 40~GeV in the polarized target fragmentation region. 

\section{Asymmetry in the reaction {\boldmath $\ppdup$} at 70~GeV.}

\begin{figure}
\vspace*{-0.5cm}
\centering
\includegraphics[width=\textwidth]{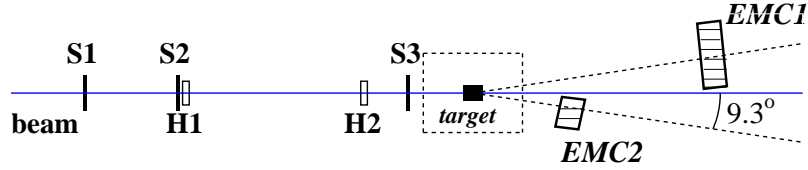}
\vspace*{-1.5cm}
\caption{Experimental setup PROZA-M: 
S1--S3 -- trigger scintillation counters; H1--H2 -- hodoscopes; 
EMC1 and EMC2 -- electromagnetic calorimeters; 
$target$ -- polarized target.}
\label{fig:setup70}
\end{figure}

$A_N$ in the reaction $\ppdup$ was measured using 70~GeV protons
extracted by a bent crystal from the accelerator vacuum chamber. 
The experimental setup is shown in Fig.~\ref{fig:setup70}. 

\noindent
Three scintillation counters S1--S3 were
used for a zero level trigger with a coincidence from two hodoscopes  
H1--H2 (each consisted of two planes).  $\gamma$-quanta were detected by two 
electromagnetic lead-glass calorimeters EMC1 and EMC2 (arrays of 480 
and 144 cells correspondingly) placed  7 and 2.8~m downstream the frozen
polarized target with 80\% average polarization. 
First level trigger on transverse energy worked 
independently for the both detectors. Angle $9.3^{\circ}$ in the laboratory
frame corresponds to $90^{\circ}$ in the center of mass system (c.m.s.)
for 70~GeV proton beam. 
We were able to detect $\pi^0$-s till $p_T \approx 3.0$~GeV/c using  
specially developed algorithm for the overlapping showers 
reconstruction \cite{lednev}. 
$\pi^0$ mass resolution was 10~MeV/$c^2$ for EMC1 and from 12 to
16~MeV/$c^2$ for EMC2. 

\begin{figure}
\centering
\begin{tabular}{cc}
\includegraphics[width=0.45\textwidth]{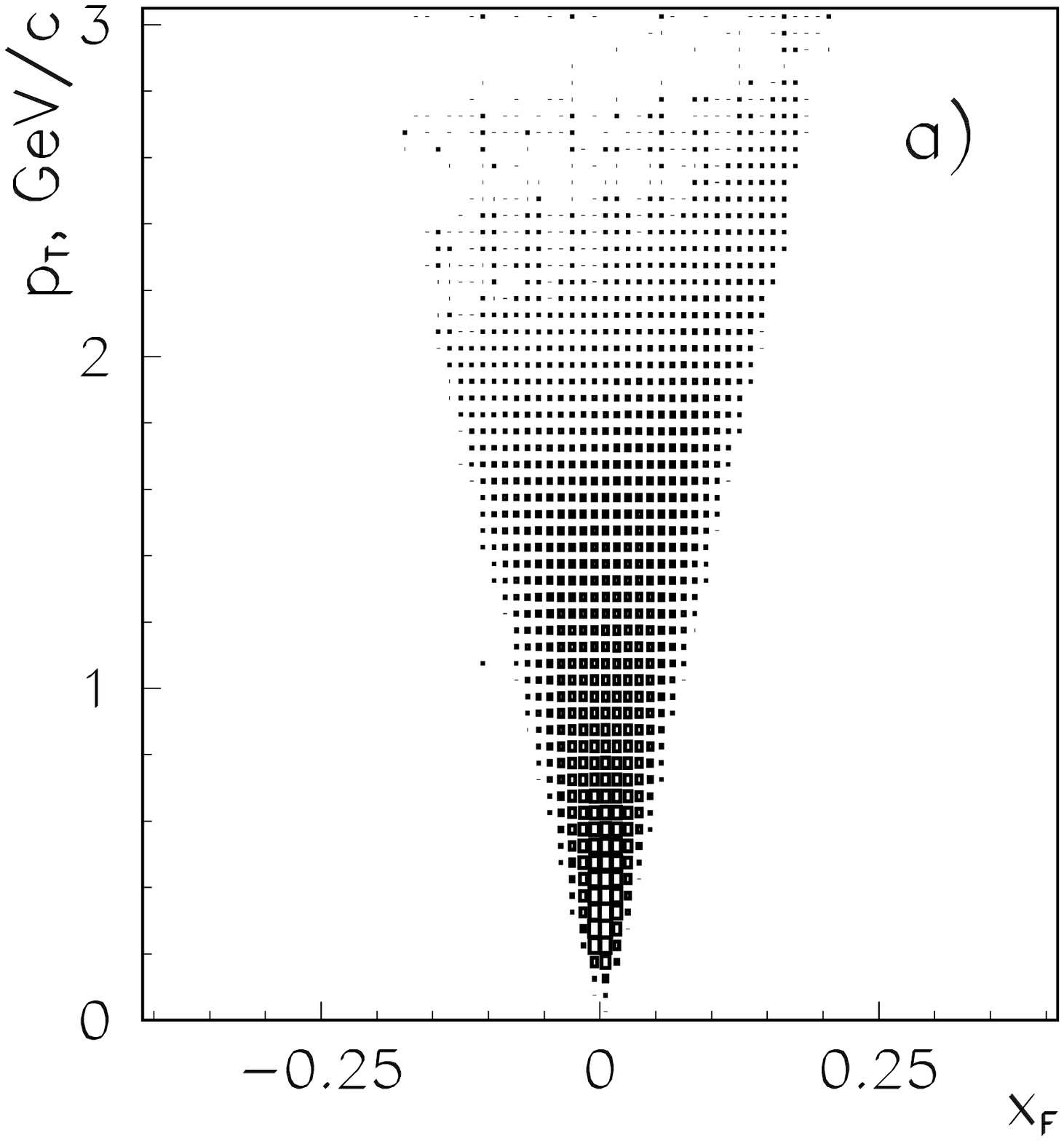} %
\includegraphics[width=0.45\textwidth]{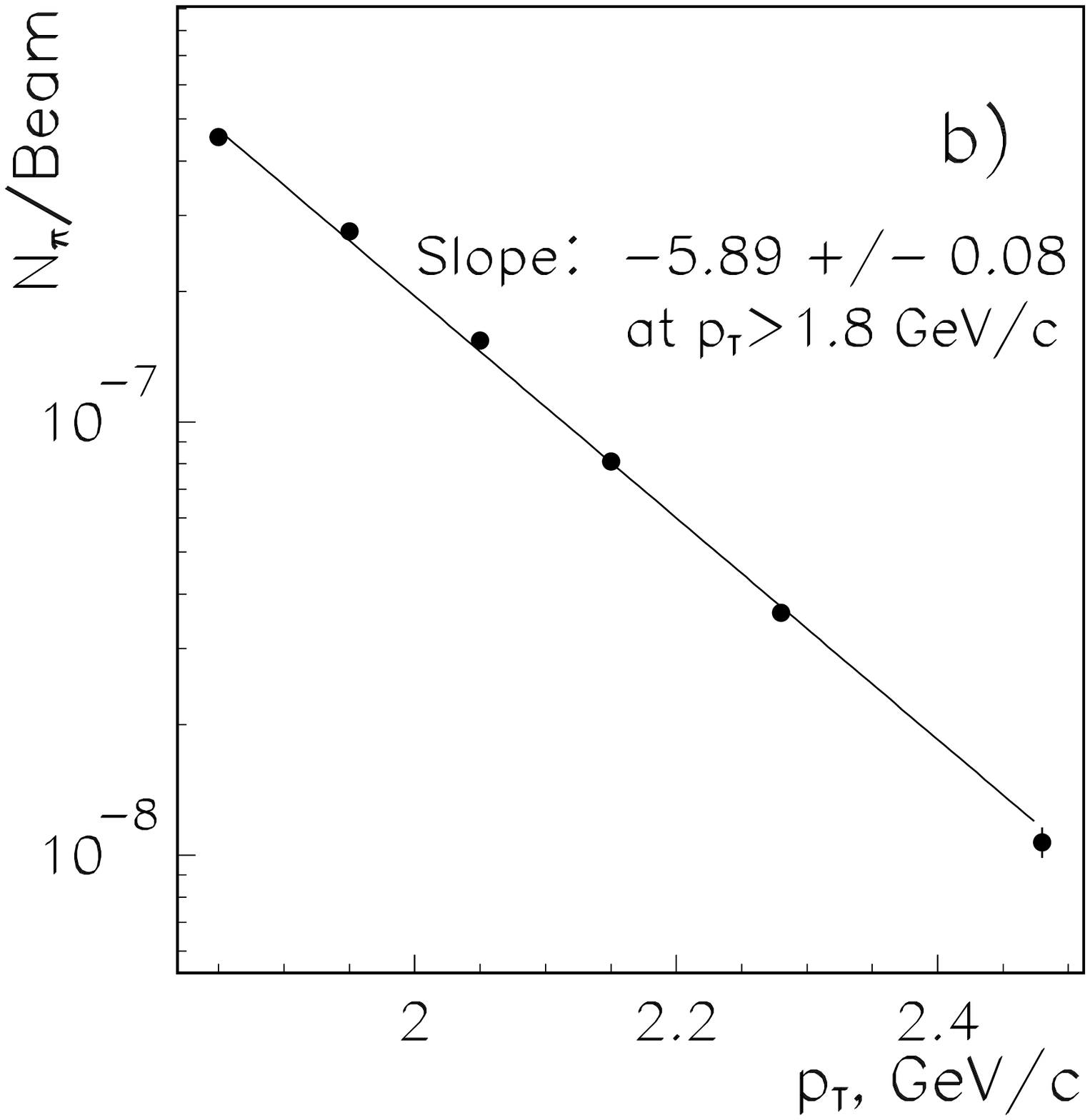} \\
\end{tabular}
\vspace*{-0.5cm}
\caption{Two-dimensional $\pi^0$ distribution on $p_T$ and $x_F$ (a) 
and relative $\pi^0$ cross-section ($N_{pions}/Beam$) (b).}
\label{fig:cross}
\end{figure}

Two dimensional distribution $(x_F,p_T)$ was symmetrical on $x_F$ 
(Fig.~\ref{fig:cross}a). The slope of relative $\pi^0$-cross-section 
presented in Fig.~\ref{fig:cross} is in good
agreement with the previous measurements of charged
pion invariant cross-section. In this experiment an exponential
constant  $\alpha=-5.89 \pm 0.08$, while FODS experiment (Protvino)
found $\alpha=-5.68 \pm 0.02$ for $\pi+$ and $\alpha=-5.88 \pm 0.02$ 
for $\pi^-$ \cite{FODScross}.  

\begin{figure}
\centering
\begin{tabular}{cc}
\includegraphics[width=0.48\textwidth,height=6.cm]{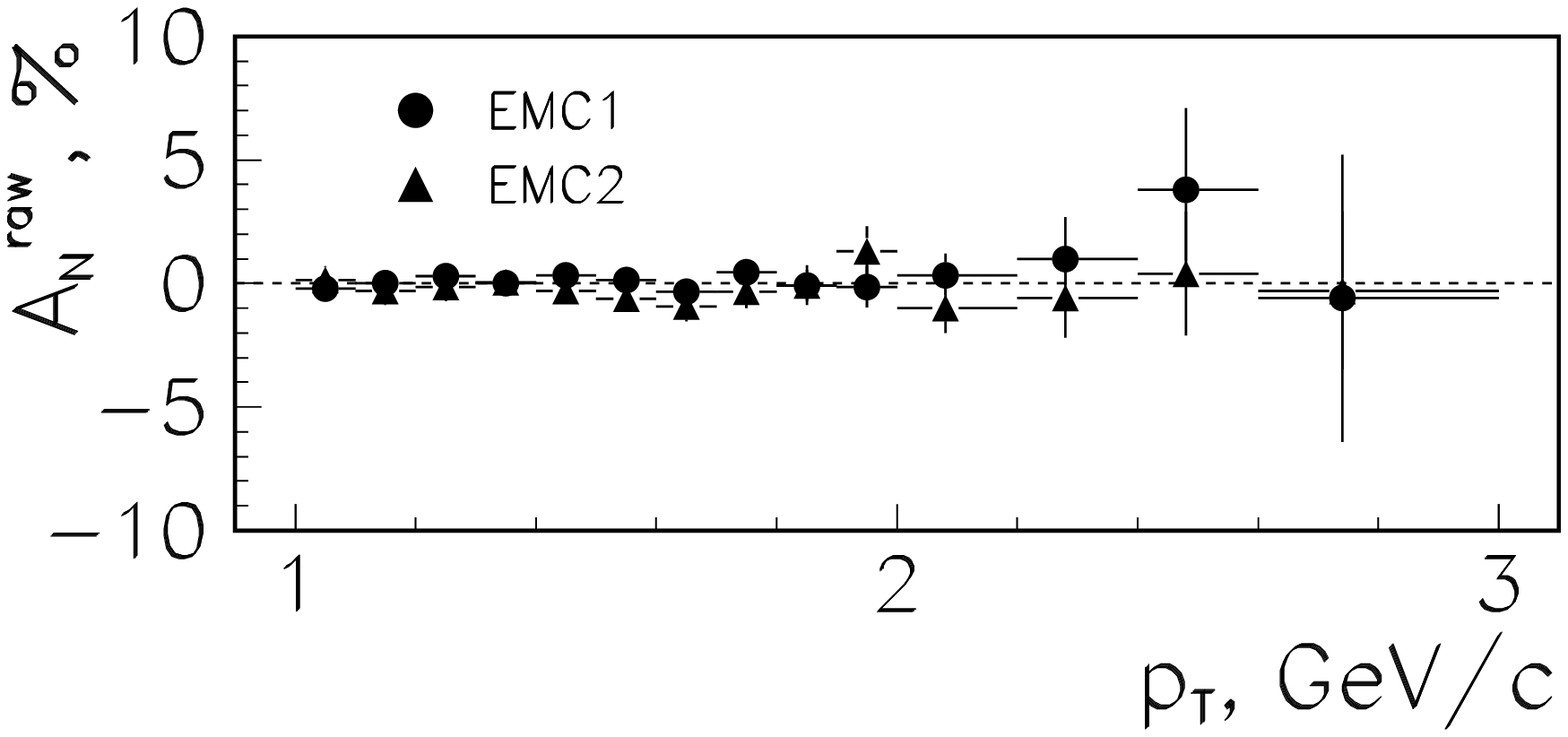} &
\includegraphics[width=0.48\textwidth,height=6.cm]{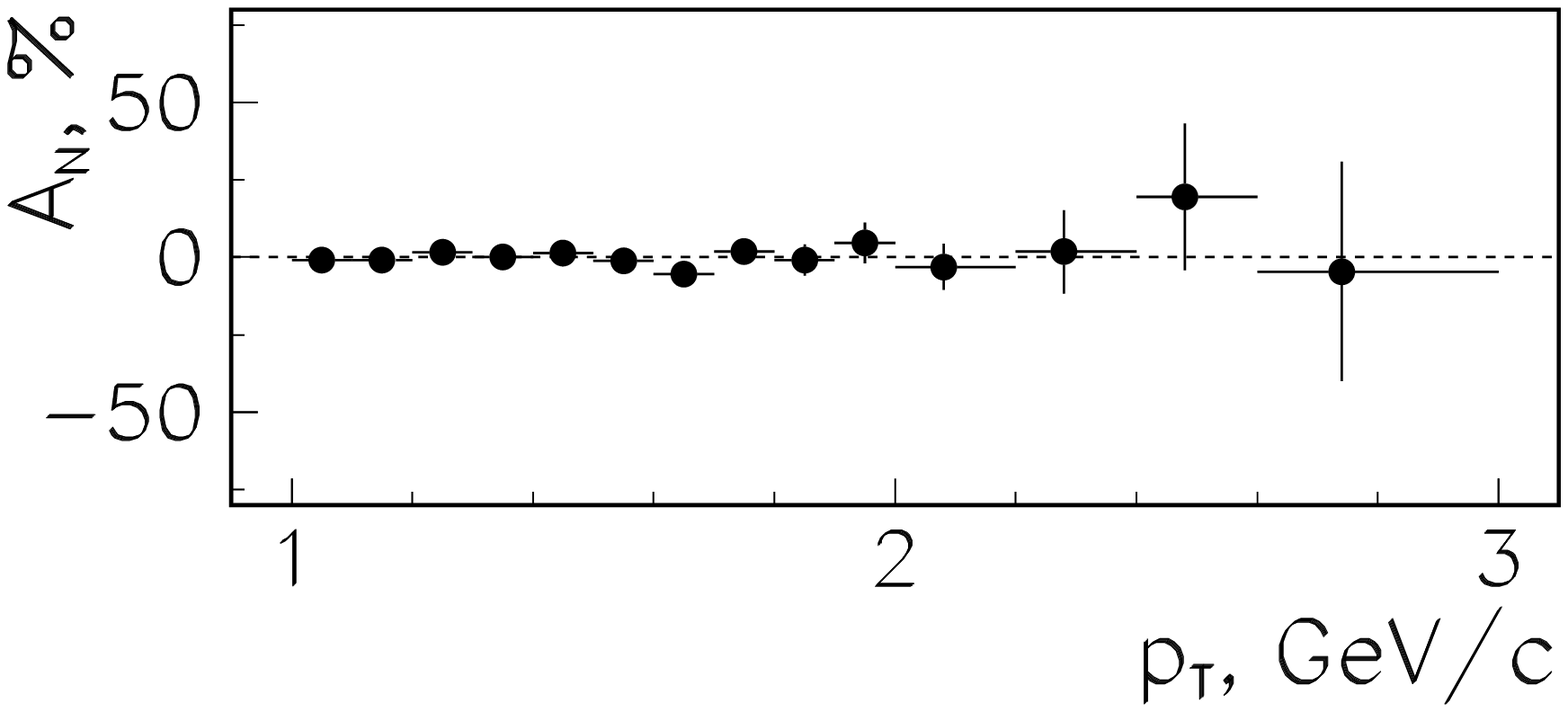}\\
\end{tabular}
\vspace*{-0.5cm}
\caption{Raw asymmetries for the two detectors separately (left) 
and summed  $A_N$ (right).}
\label{fig:proza70_asym}
\end{figure}

The asymmetry $A_N$ with a detector to the right of the beam 
line is defined as
\centerline{\large $A_N=\frac{D}{P_{target}}\cdot A_N^{raw} =
\frac{D}{P_{target}}\cdot 
\frac{n_{\downarrow}-n_{\uparrow}}{n_{\downarrow}+n_{\uparrow}}$,}

\noindent where $D$ --- target dilution factor,
$n_{\downarrow}$ and $n_{\uparrow}$ --- normalized number of 
$\pi^0$-mesons for opposite target polarizations.
False asymmetry was investigated by dividing statistics with the
same target polarization on two data samples 
and calculating the asymmetry for them. 
False asymmetry is close to zero. We compared the
asymmetry for the two detectors and did not find any
difference (see Fig.~\ref{fig:proza70_asym}, left).

The final result for the both detectors is presented in 
Fig.~\ref{fig:proza70_asym}  (right). The asymmetry is
zero within the error bars. Our result is in agreement with the
FNAL data at 200~GeV \cite{e704cent} and contradicts to the previous
CERN measurements at 24~GeV \cite{cern}. Comparing the presented data 
with the $\pi^0$ asymmetry $A_N \approx -40\%$ at 40~GeV \cite{protv},  
we may conclude that the asymmetry depends on quark flavour. 
Otherwise we have to suppose significant changes in interaction dynamics
in the energy range between 40 and 70~GeV.

\section{ {\boldmath $A_N$} in the reaction {\boldmath $\pimp$} at 40~GeV.}

\begin{figure}
\centering
\begin{tabular}{cc}
\includegraphics[width=0.48\textwidth]{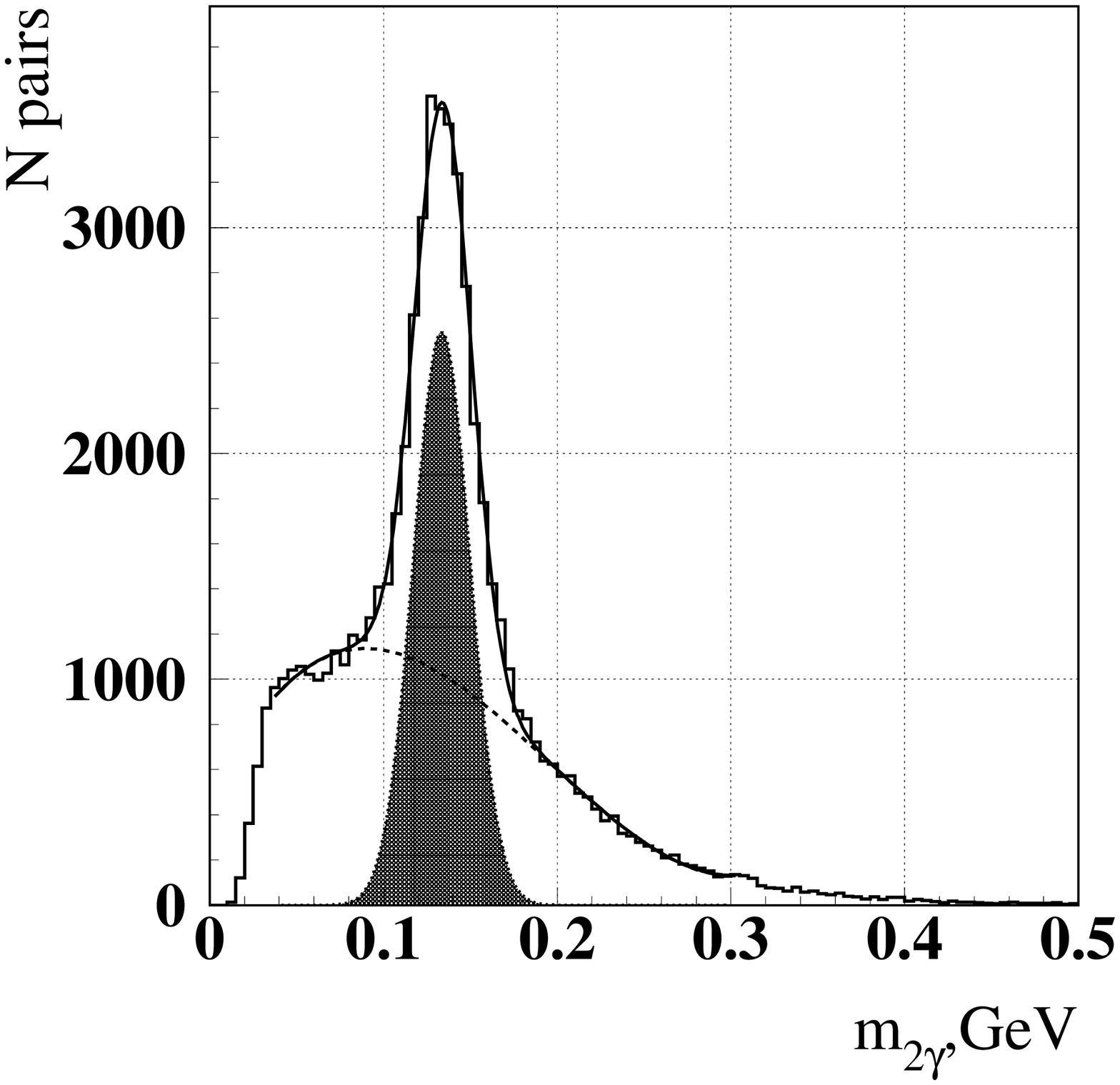} &
\includegraphics[width=0.48\textwidth]{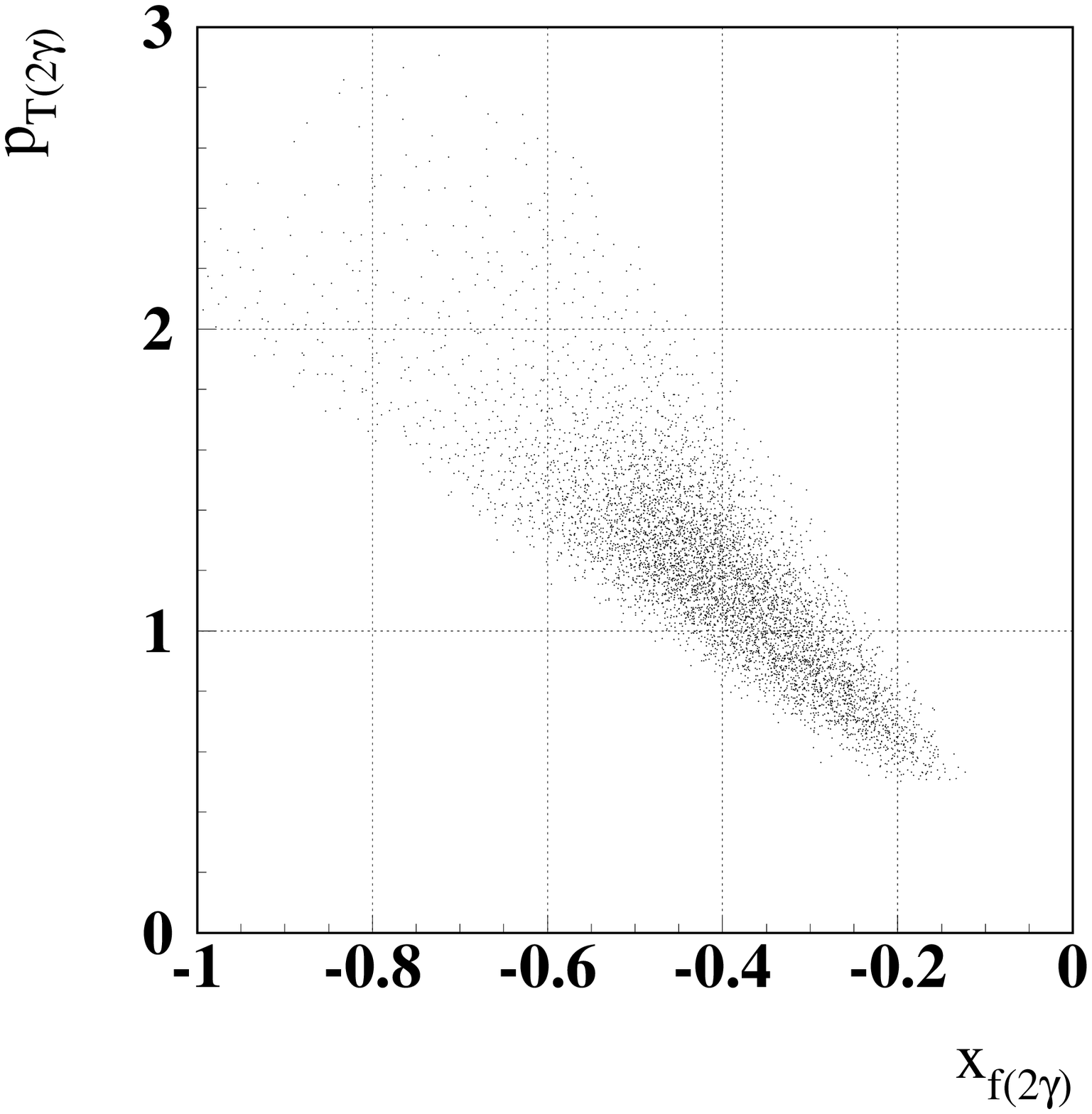} \\
\end{tabular}
\vspace*{-0.5cm}
\caption{Mass spectrum (left) and two-dimensional distribution on 
kinematic variables for the reaction $\pimp$ at 40~GeV.}
\label{fig:mass40}
\end{figure}

The measurements in the reaction $\pimp$ at 40~GeV were done in 
2000 with the modified experimental setup PROZA-M.
The electromagnetic calorimeter of 720 cells was placed at 2.3~m downstream
the target at the angle of $40^{\circ}$ in the laboratory frame to measure
the asymmetry in the polarized target fragmentation region. A mass spectrum
and the detector kinematic region are presented in Fig.~\ref{fig:mass40}. 

\begin{figure}
\begin{tabular}{cc}
\includegraphics[width=0.35\textwidth,height=4.5cm]
{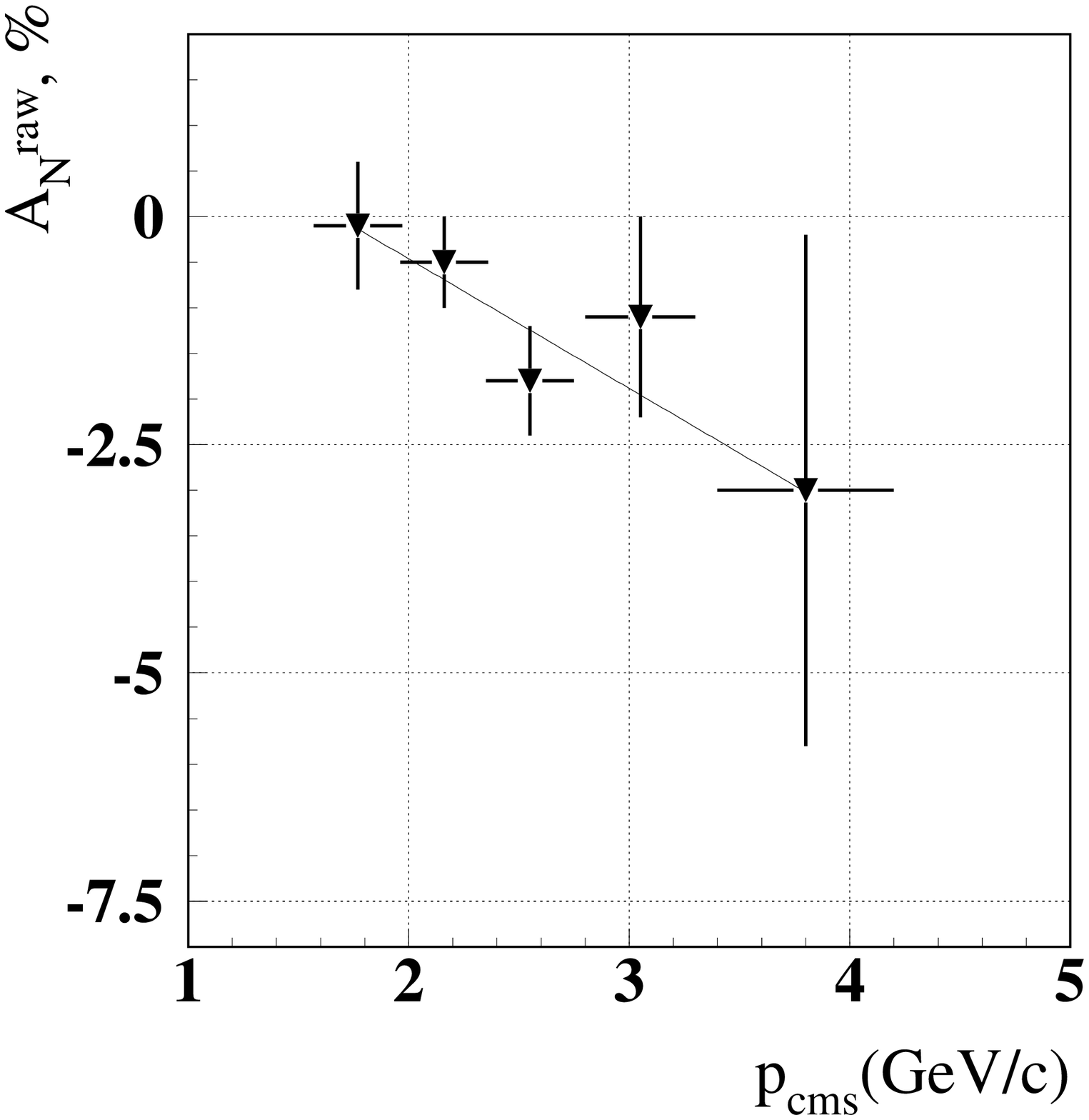} &
\includegraphics[width=0.65\textwidth,height=4.5cm]
{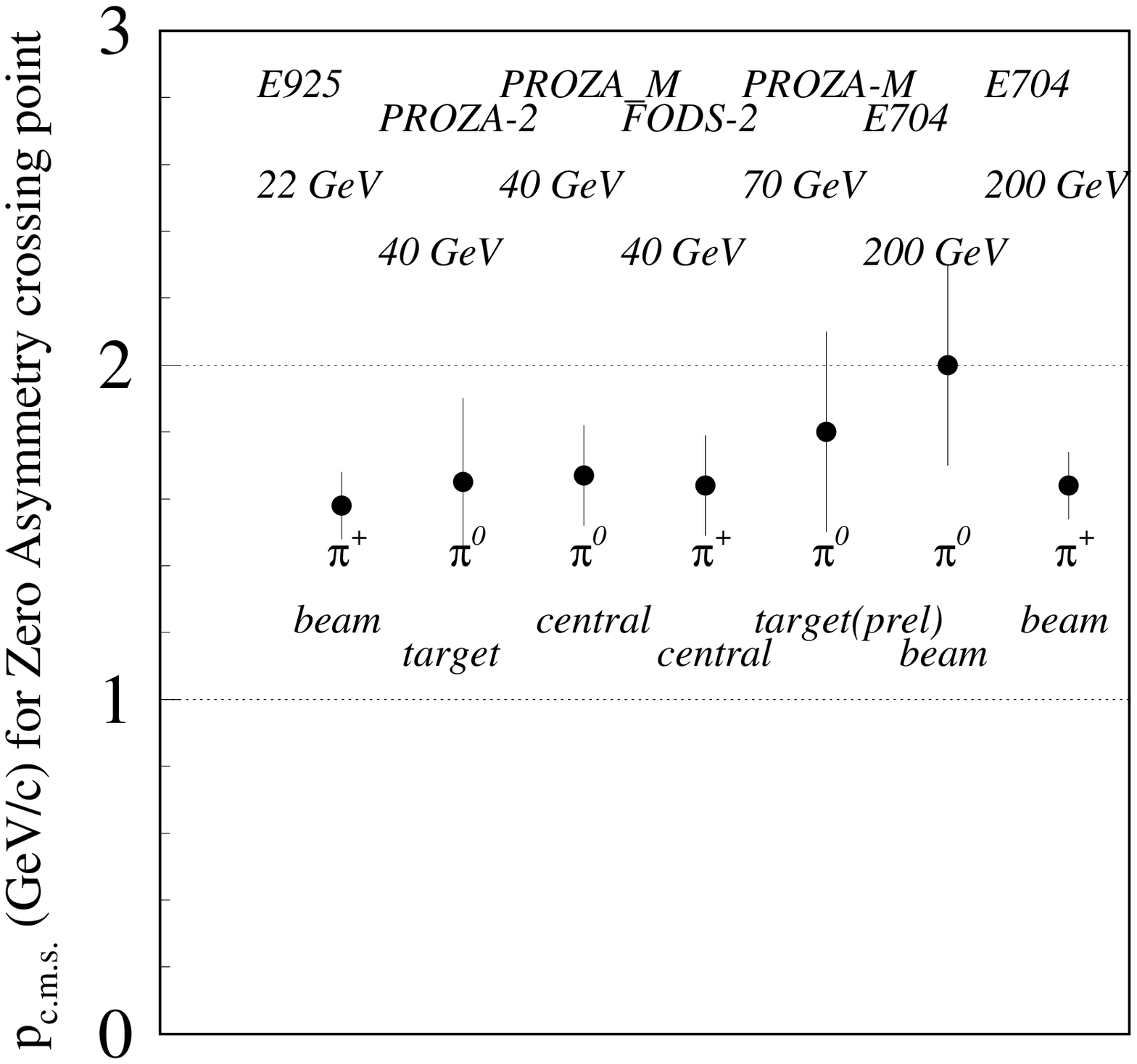}\\
\end{tabular}
\vspace*{-0.5cm}
\caption{The dependence of $A_N$ on momentum in c.m.s. (left) and 
center of mass momentum values where  the pion asymmetry 
starts to grow up for different experiments.}
\label{fig:scal}
\end{figure}

The measured asymmetry $A_N$ is close to zero at low values of
$|x_F|$ and $A_N=(-15 \pm 4)\%$ at $-0.8<x_F<-0.4$. The result is 
similar to the $\pi^0$ asymmetry in the polarized beam fragmentation 
experiments E704 ({($12.4\pm 1.4)\%$}, $\sqrt s=20$~GeV \cite{e704beam})
and STAR ({$(14\pm 4)\%$, $\sqrt s=200$~GeV \cite{STAR2002}}). 
Earlier we also measured the asymmetry in this reaction in the central region 
and  found that $A_N$ starts to rise up at 
$p_T \approx 1.6$~GeV/c \cite{protv}.  We studied the asymmetry 
dependence on c.m.s. momentum and surprisingly found 
the asymmetry to begin to grow up at the same momentum 
$p_T^0 (cms) = 1.70 \pm 0.25$~GeV/c (Fig.~\ref{fig:scal}, left).

\begin{figure}
\centering
\includegraphics[width=0.8\textwidth,height=5.cm]{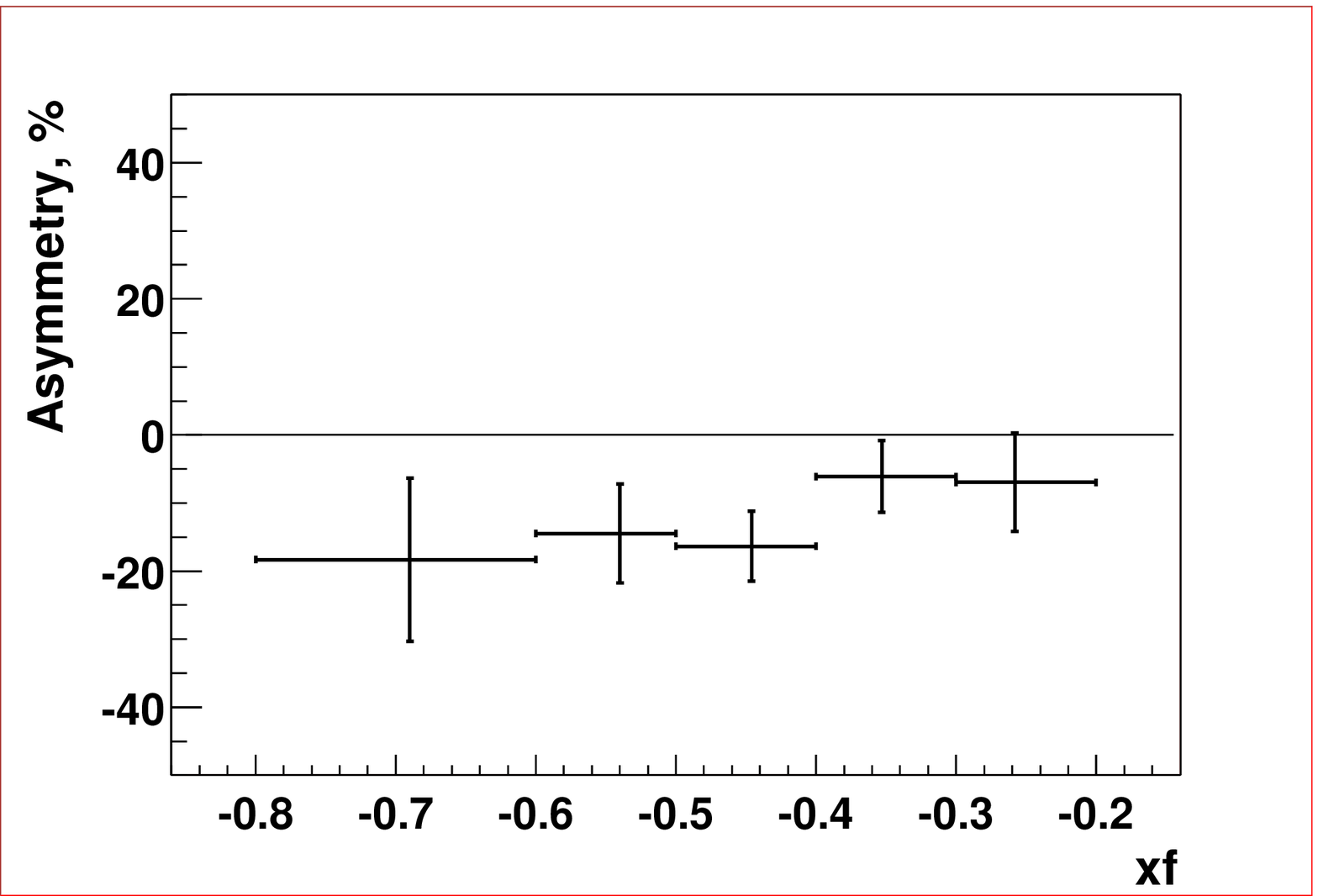}
\vspace*{-0.5cm}
\caption{Asymmetry $A_N$ in reaction $\pimp$ at 40~GeV.}
\label{fig:asym}
\end{figure}  

We have analysed the experimental data of other experiments and concluded
that the asymmetry arises in the energy range between
1.5 and 2.0~GeV for all fixed target experiments (Fig.~\ref{fig:scal}, right).

\section*{Conclusions}

Finally we may summarize:
\begin{itemize}
\item $A_N$ in  the reaction { $p + p_{\uparrow} \rightarrow \pi^0 + X$} 
at 70~GeV equals to zero in the central region for $1.0<p_T<3.0$~GeV/c 
and is in agreement with the E704 result.
\vspace*{-0.2cm}
\item  In the reaction { $\pimp$} at 40~GeV in the polarized 
target fragmentation region $A_N=(-15 \pm 4)\% $ at 
$-0.8<x_F<-0.4$ and $p_T>0.8$~GeV/c  and close to zero at small $|x_F|$ 
and $p_T<1.5$~GeV/c.
\vspace*{-0.2cm}
\item The last result is similar to the $\pi^0$ asymmetry in the polarized 
beam fragmentation region measured by E704 and STAR experiments.  
The $\pi^0$ inclusive production in the polarized proton fragmentation
region can be considered as a proper reaction for polarimetry.

\vspace*{-0.2cm}
\item The asymmetry in fixed target experiments arises at  
$p_{cms}$ from 1.5 to 2.0~GeV/c independently on the beam energy and 
kinematic region.
\end{itemize}

{\it The work is supported by RFBR grant 03-02-16919}

\lastevenpage
\end{document}